\documentclass[prb,twocolumn,groupeaddress]{revtex4-2}

\usepackage{amsmath}
\usepackage{graphicx}
\usepackage[colorlinks=true, allcolors=blue]{hyperref}

\begin{document}
\title{Quasiparticle approach to the transport in infinite-layer nickelates}
  \author{Steffen B\"otzel}
  \affiliation{Institut f\"ur Theoretische Physik III, Ruhr-Universit\"at Bochum,
  D-44780 Bochum, Germany}
\author{Ilya M. Eremin}
\affiliation{Institut f\"ur Theoretische Physik III, Ruhr-Universit\"at Bochum,
  D-44780 Bochum, Germany}
\author{Frank Lechermann}
\affiliation{Institut f\"ur Theoretische Physik III, Ruhr-Universit\"at Bochum,
  D-44780 Bochum, Germany}

\begin{abstract}
 The normal-state transport properties of superconducting infinite-layer nickelates are investigated within an interacting three-orbital model. It includes effective Ni-$d_{z^2}$, Ni-$d_{x^2-y^2}$  bands as well as the self-doping band degree of freedom. Thermopower, Hall coefficient and optical conductivity are modelled within a quasiparticle approximation to the electronic states. Qualitative agreement in comparison to experimentally available Hall data is achieved, with notably a temperature-dependent sign change of the Hall coefficient for larger hole doping $x$. The Seebeck coefficient changes from negative to positive in a non-trivial way with $x$, but generally shows only modest temperature dependence. The optical conductivity shows a pronounced Drude response and a prominent peak structure at higher frequencies due to interband transitions. While the quasiparticle picture is surely approximative to low-valence nickelates, it provides enlightening insights into the multiorbital nature of these challenging systems.
\end{abstract}

\maketitle

\section{Introduction}
Thin films of low-valence nickelates in infinite-layer and multilayer form show superconductivity upon hole doping with a $T_{\rm c}\sim 15$\,K~\cite{li19,li20,zeng20,pan21}. These long sought-after findings brought new life to the research on superconducting oxides and challenge also the current understanding of high-$T_{\rm c}$ layered cuprates. In fact, while the superconducting domes with hole doping in these structurally akin nickelates and cuprates resemble each other~\cite{li20}, various normal-state properties apparently differ. For instance, though sizable antiferromagnetic (AFM) correlations appear present in infinite-layer nickelates RENiO$_2$~\cite{hayward99,lu20,cui2021}, with rare-earth ion RE=La,Pr,Nd, the solid stoichiometric AFM order known from cuprates remains elusive. 

Furthermore, the transport properties of RENiO$_2$, with and without hole doping of the form RE$_{1-x}$Sr$_x$NiO$_2$, display remarkable differences to the cuprates ones. At stoichiometry, the (charge-transfer) insulating character of copper oxides is contrasted by weak metallicity in the nickel oxides. This experimental finding of a system inbetween metal and insulator has been foreseen by an early quantum-chemistry description by Choisnet {\sl et al.}~\cite{choisnet96}. Later density functional theory (DFT) calculations~\cite{anisimov99,lee04,nomura19,lechermann20-1,olevano20} proved the existence of an additional so-called self-doping (SD) band that adds electron pockets to an otherwise Ni-$d_{x^2-y^2}$ dominated Fermi surface in the weakly correlated limit. These electron pockets are the result of hybridization between non-$d_{x^2-y^2}$ Ni$(3d)$ and RE$(5d)$ orbitals. First principles calculations show that the electron pockets remain at the Fermi level upon including electron correlation beyond DFT. Yet the degree of quasiparticle (QP) character of the original Ni-$d_{x^2-y^2}$ band is a matter of debate from that vantage 
point~\cite{chen22}. With hole doping $x$, the resistivity remains sizable in the well-underdoped regime $x<0.12$ and also in the well-overdoped regime $x>0.25$. Thus, the charge-carrier transport is not a straightforward metallic one. Although recent improvements in thin-film preparation yield an increase in conductivity~\cite{lee22}, it does not quite match the wide-temperature-range Fermi-liquid signature of overdoped cuprates. Near optimal doping, the normal-state transport is linear in temperature~\cite{lee22}, resembling cuprate transport in that regime. Note also that the Hall coefficient in nickelates changes sign from negative to positive in the superconducting doping region~\cite{li20}, a feature which in experimental works is modelled by an effective two-band picture.

Theoretical accounts of hole-doped infinite-layer nickelates do agree on the principle
fact that the SD electron pocket around the $\Gamma$ point in reciprocal space is shifted upwards in energy and away from Fermi level. But the precise doping level $x$ where this happens depends on the specific theoretical method (setting) used. Note that angle-resolved photoemission measurements providing an identification of the near-Fermi-level electronic states are so far lacking. A concrete picture of the low-energy landscape has for now to rely solely on theoretical pictures. This especially concerns the relevant Ni$(3d)$ orbitals upon changing $x$. Standard DFT and DFT+dynamical mean-field theory (DMFT) assessments mark the sole relevance of the Ni-$d_{x^2-y^2}$ dispersion, possibly with some remaining SD-pocket contribution, as the key to understand the low-lying electronic states~\cite{wu19,zhangzhang20,karp20,leonov20,adhikary20,been21,geisler21}. Approaches that additionally allow for explicit ligand-based correlations designate the competition between Ni-$d_{x^2-y^2}$ and Ni-$d_{z^2}$ as essential to decypher the low-energy processes~\cite{lechermann20-1,lechermann2020multiorbital,petocchi20}. There are further approaches, e.g. favoring Hunds physics~\cite{werner20,kang21} or a description based on Ni-$d_{x^2-y^2}$ and some effective (interstitial) orbital degree of freedom~\cite{gu20,plienbumrung22,jiang22}.

In this work, we study the transport in the normal state of infinite-layer nickelates based on the theoretical picture of additionally relevant ligand-based correlations. More concretely, we utilize an interacting model Hamiltonian tailored to the low-energy part of a comprehensive first-principles many-body description of stoichiometric and hole-doped NdNiO$_2$~\cite{lechermann2020multiorbital}.
The latter is given by a combination of DFT, DMFT with a Ni-based
correlated subspace, and including explicit Coulomb interaction on oxygen via the self-interaction correction (SIC), the so-called DFT+sicDMFT approach (see Ref.~\onlinecite{lechermann19} for more details). Within a quasiparticle approximation to the electronic states, we here reveal key features of the doping-dependent thermopower, Hall coefficient and optical conductivity.

\section{Theoretical Approach}
To calculate the transport properties for the normal state of Nd$_{1-x}$Sr$_{x}$NiO$_2$ for different hole dopings we use a three-dimensional low-energy description of the renormalized band structure. In Ref.~\onlinecite{lechermann2020multiorbital} it is argued that three effective orbitals are sufficient to describe the electronic degrees of freedom near the Fermi level, namely the Ni-$d_{z^2}$ and Ni-$d_{x^2-y^2}$ orbitals as well as an effective orbital giving rise to the SD band. The latter contains weights from Nd$(5d)$ orbitals at low energy merged with weights from the remaining Ni-$t_{2g}$, Ni$(4s)$ and O$(2p)$ orbitals.

The model Hamiltonian is obtained from maximally-localized Wannier~\cite{marzari12} downfolding of the DFT band structure to obtain the hopping integrals $t^{mm'}_{ij}$ between the aforementioned orbitals denoted with $m,m'$ and sites $i,j$. Adding interaction terms, the Hamiltonian reads
\begin{align}
    H = \sum_{i\neq j,mm'\sigma} c^{\dagger}_{im\sigma} c_{jm'\sigma} + \sum_{i} \left( H_i^\text{int} + H_i^{\text{orb}} \right),
\end{align}
where $\sigma$ denotes the spin projection. The part $H_i^\text{int}$ describes interactions between both Ni-$e_g$ orbitals via a Slater-Kanamori form, whereas the $H_i^{\text{orb}}$ term allows for a shift of the SD band, corrects for double-counting terms of the Ni-$e_g$ interaction and includes crystal-field terms~\cite{lechermann2020multiorbital}. This minimal Hamiltonian is then solved with rotationally invariant slave bosons at saddle point~\cite{lechermann2007rotationally}. At strong coupling, the model resembles the physics obtained from the more generic DFT+sicDMFT approach to infinite layer nickelates. The two key features of this physics are the (near) orbital-selective Mott-insulating state of Ni-$d_{x^2-y^2}$ and a doping-dependent shift of the occupied Ni-$d_{z^2}$ dispersion branch in the $k_z=1/2$ part of the Brillouin zone, starting to cross the Fermi level at $x~\sim 0.1$. Note that in the multi-orbital slave-boson description, the quasiparticle (QP) weight $Z$ for the Ni-$d_{x^2-y^2}$ dispersion also remains finite, but below $Z=0.2$ for all hole dopings. In the following we will denote the slave-boson renormalized form of the Hamiltonian $H$ as $\Tilde{H}$. The latter Hamiltonian has recently been employed to compute the resonant inelastic x-ray spectroscopy (RIXS) spectrum of spin fluctuations and resulting
superconducting instabilities~\cite{kreisel22}. Here it will be used to elucidate a minimal and simple perspective on the intriguing transport properties of infinite-layer nickelates.

The Kubo formalism for correlated electrons~\cite{jarrell95,oudovenko06,arita08,tomczak2009optical,boehnke14} is applied to the renormalized Hamiltonian using the dipole approximation, so that we compute it in the long-wavelength limit $\mathbf{q}=0$. Furthermore, vertex corrections are neglected. These approximations lead to the following expression for the real part of the frequency-dependent electric conductivity
\begin{equation}
	\sigma^{\nu\nu}(\omega) =
	\frac{2e^2\pi}{\hbar\Omega} 
	 \int\text{d}\omega'\, 
	\frac{f(\omega')-f(\omega+\omega')}{\omega}\,\tau^{\nu\nu}(\omega')
 \label{Eq:OptCond}
\end{equation}
using
\begin{equation}
	\tau^{\nu\nu}(\omega) =\frac{1}{N}	\sum_{\mathbf{k}} \text{Tr} \left[ A_\mathbf{k}(\omega+\omega')v_\mathbf{k}^\nu A_\mathbf{k}(\omega')v_\mathbf{k}^\nu \right]\;,
 \label{Eq:kernel}
\end{equation}
where $\nu$ is an index for the renormalized band. Here the trace is performed in orbital space and the sum over the spin indices is assumed. The spectral function is given by
\begin{equation}
A^{mm'}_\mathbf{k}(\omega) = -\frac{1}{\pi} Z_{mm'}\,\text{Im} \left( \omega - \Tilde{H}_{\mathbf{k}} - i\gamma \right)_{mm'} ^{-1}
\end{equation}
with the scattering rate $\gamma$ chosen to be energy-independent, corresponding to a dominant elastic part. This assumption has already been used to account for the transport in reduced five- and three-layer nickelates~\cite{grissonnanche2022}. In addition, the high level of disorder causes a large elastic scattering rate for which the Seebeck and Hall coefficients are insensitive to details of its the temperature dependence.  Further, $\Omega$, $N$ and $f(\omega)$ denote the volume of the unit cell, the number of points in ${\bf k}$-space and the Fermi function, respectively. The components of the Fermi velocities with explicit orbital dependence are given by
\begin{equation}
	v^{\nu,mm'}_{\mathbf{k}} = 
	\nabla_{{\bf k}^{\nu}} H_{\mathbf{k}}^{mm'} - i(\rho^{\nu}_{m}-\rho^{\nu}_{m'}) H_{\mathbf{k}}^{mm'}\;.
\end{equation}
The second term, in which $\rho^{\nu}_{m}$ is the displacement of the Wannier orbitals with respect to the center of the unit cell, is important for multi-atomic unit cells~\cite{tomczak2009optical}. 

Applying the limit $\omega\rightarrow 0$ in Eq.~(\ref{Eq:OptCond}) yields the DC conductivity 
\begin{equation}
	\sigma^{\nu\nu} = \lim_{\omega\rightarrow 0} \sigma^{\nu\nu}(\omega) = 
	\frac{2e^2\pi}{\hbar \Omega} 
	\int \text{d}\omega'\, 
	\left( -\frac{\text{d}f}{\text{d}\omega'} \right)  \tau^{\nu\nu}(\omega=0)\;.
 \label{Eq:currentcurrent}
\end{equation}
Analogously, in the same limit the current-heat correlation function reads
\begin{equation}
	\alpha^{\nu\nu} =
	\frac{2e^2\pi}{\hbar \Omega} 
	\int \text{d}\omega'\, 
	\left( -\frac{\text{d}f}{\text{d}\omega'} \right) \omega'\, \tau^{\nu\nu}(\omega=0)\;.
 \label{Eq:current-heat}
\end{equation}
The ratio of the traces of the current-heat and current-current correlation functions yields the Seebeck coefficient
\begin{equation}
    S = -\frac{k_B}{|e|} \frac{\alpha}{\sigma}\;.
\end{equation}
The additional $\omega'$ in the frequency integral of Eq.~(\ref{Eq:current-heat}) weights contributions from above and below the Fermi surface with opposite sign such that states above (below) lead to negative (positive) contributions to $S$. For pure elastic scattering, the Seebeck coefficient is thus a measure of particle-hole-asymmetry of the electronic dispersion.

Besides the Seebeck coefficient and the frequency-dependent conductivity, we also obtain a simple estimate for the Hall coefficient 
\begin{equation}
    R_\text{H} = \frac{1}{|e|}\,\frac{\sigma_{xy}}{B\sigma_{xx}^2}\;,
    \end{equation}
with $B$ is the magnetic field pointing along $z$-direction. Calculating the real part of the Hall conductivity $\sigma_{xy}$ is more intricate than for the diagonal components \cite{markov2019robustness} and performing the required traces in three spatial dimensions is numerically costly. Therefore, to get a first approximation, we neglect interband terms and calculate contributions from the three-band model following Refs.~\onlinecite{kuchinskii2022hall,voruganti1992conductivity}, i.e.   
\begin{align}
	\sigma_{xy} = & \frac{2\pi^2e^3B}{3\hbar^2\Omega}\int
 \text{d}\omega\,\frac{\text{d}f}{\text{d}\omega}\times \nonumber \\
	& \frac{1}{N}	\sum_{\bf k}  A_\mathbf{k}(\omega)^3
	\left(  \frac{\partial\epsilon_\mathbf{k}}{\partial k_x} \right)^2 
	\frac{\partial^2\epsilon_\mathbf{k}}{\partial k_x^2}\;.
\end{align}
We have checked for several settings that using this approach and using the rigorous treatment in orbital space described above give similar qualitative results for the Seebeck coefficient despite some minor quantitative differences.

Appropriate convergence of the k integration over the Brillouin zone, for example according in Eq.~(\ref{Eq:kernel}), is ensured by the choice of the k-mesh. Explicitly, we have used at least $540^3$ points for the Seebeck and at least $600^3$ points for 1/8 of the Brillouin zone for the calculation of the Hall coefficient. For the optical conductivity sufficient convergence is achieved with $120^3$ k-points. Numerical integration over energies is always done with 100 points. Due to the appearance of $df/d\omega$ in Eqs.~(\ref{Eq:currentcurrent}-\ref{Eq:current-heat}), which is strongly peaked at the Fermi energy, an energy cut-off of $\pm5k_BT$ is used, where $df/d\omega$ drops to less than three percent of its peak value. 

Before discussing the results, let us generally remind about the QP approximation taken in this work. For instance, in a recent doping-dependent DFT+sicDMFT study~\cite{lechermann21} it was revealed that especially in the doping region between stoichiometry and the superconducting phase, the electronic spectrum turns out very incoherent. In other word, well-defined QPs are only expected in certain regions of the infinite-layer nickelate phase diagram. Still, we believe that a QP-based transport description should be performed and weighed as a reference for further studies. In addition, the sole impact of band-dependent features, dressed by proper renormalization from electron-electron interaction, may be investigated ideally in this limit.

\section{Results}
\begin{figure}[h]
	\begin{center}
      \includegraphics[width=\linewidth]{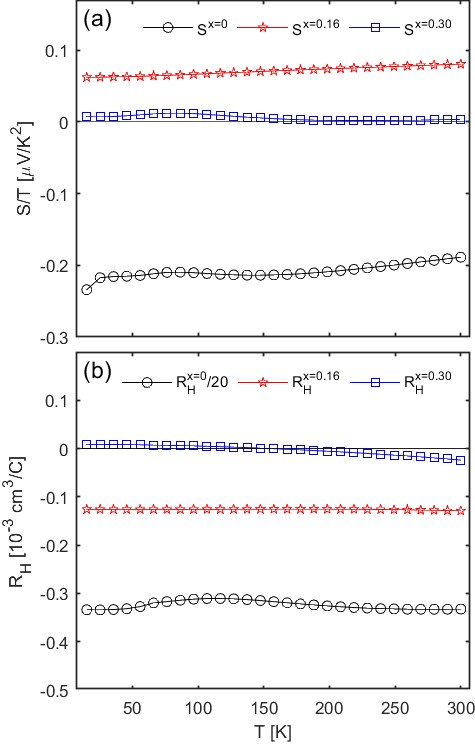}
	\end{center}
	\caption{Calculated temperature dependence of the Seebeck coefficient (a) and the Hall coefficient (b) for various doping concentrations $x$. Note, the Seebeck coefficient is small but remains positive for $x = 0.3$. } \label{Figure1}
\end{figure}

\begin{figure*}[t]
	\begin{center}
		\includegraphics[width=\linewidth]{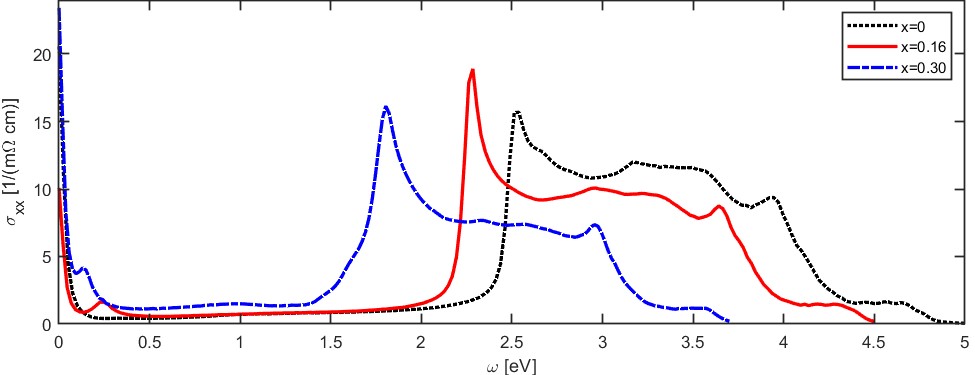}
	\end{center}
	\caption{Calculated optical conductivity for various dopings at a temperature $T = 300$ K.} \label{Figure2}
\end{figure*}

\begin{figure*}[t]
	\begin{center}
		\includegraphics[width=\linewidth]{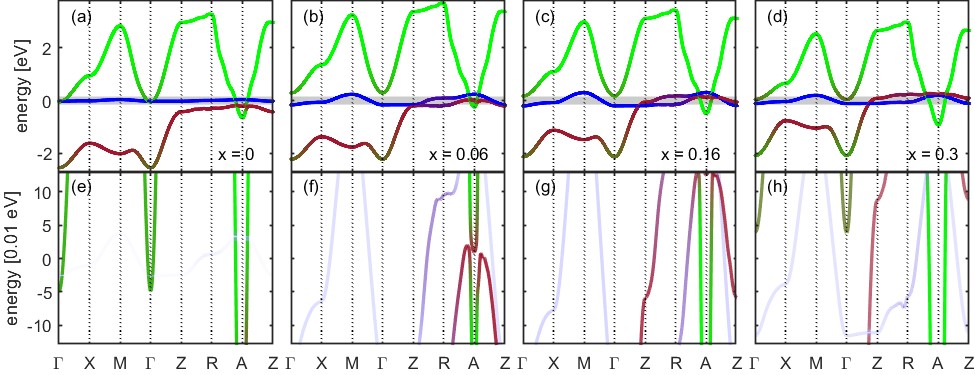}
	\end{center}
	\caption{Calculated renormalized electronic band structure shown along the high-symmetry points $\Gamma$-X-M-$\Gamma$-Z-R-A-Z. The orbital weights are shown by different colors. In particular, dark red, blue, and green correspond to the Ni-$d_{z^2}$, Ni-$d_{x^2-y^2}$ and the SD-orbital, respectively. The thin grey boxes in panels \textbf{(a)}-\textbf{(d)} highlight the energy window $5\,k_BT$ for $T = 300K$, relevant for the DC conductivity. Panels \textbf{(e)}-\textbf{(f)} show the corresponding zoom in to that energy window. In addition, the shading of the colors follows the QP weight, $Z$, of the corresponding bands, shown in the lower panel. This mainly affects the (nearly) half-filled Ni-$d_{x^2-y^2}$ band with small $Z<0.2$.}
   \label{Figure3}
\end{figure*}

In what follows we present the results for the Seebeck coefficient (Fig.~\ref{Figure1}(a)), the Hall coefficient (Fig.~\ref{Figure1}(b)) and the optical conductivity (Fig.~\ref{Figure2}) based on the effective three-band model $\tilde{H}$ calculated for three characteristic hole concentrations $x=0$, $x=0.16$ and $x=0.3$. To better understand the behaviors we also plot the renormalized band structure in Fig.~\ref{Figure3}.


As seen from Fig.~\ref{Figure1}(a) the Seebeck coefficient is weakly temperature dependent for all shown dopings, but negative at stoichiometric concentration and positive for $x=0.16,0.3$. This evolution is clearly connected to the upwards shift of the Ni-$d_{z^2}$ band from below to above the Fermi level with increasing hole concentration (see Fig.~\ref{Figure3} from left to right). At stoichiometry it is not inside the range $\pm5\,k_BT$ for temperatures below 300\,K (panel Fig.~\ref{Figure3}(e)). Since Ni-$d_{x^2-y^2}$ is quasi-localized and the corresponding QP band very flat, all relevant contributions must stem from the SD band, which has electron character at the $\Gamma$ point and A point (see Fig.~\ref{Figure3}(a)). 

In the present effective three-band picture, the SD band is also found to be explicitly important for higher dopings. Note also that at $x=0.06$ the avoided crossing of the SD band and the Ni-$d_{z^2}$ band leads to hole-band like features directly at the Fermi level around the A point of the BZ, as shown in panel Fig.~\ref{Figure3}(f). These features would result in a larger and temperature-dependent positive Seebeck coefficient. 
However, so far it has been difficult to investigate this very low doping range experimentally in great detail, and as mentioned above, it may be that in reality the QP picture breaks down in this region. Because of this uncertainty, we focus here on somewhat higher dopings where the avoided crossings are shifted upwards. For $x=0.16$ the particle-hole asymmetry of the dispersion at the Fermi level turns out to be positive and shrinking with temperature. For $x=0.3$ this trend continues and results in tiny, but positive values for all temperatures. The upper band is getting close to the Fermi surface at $\Gamma$ and contributes to the thermopower with negative sign. However, these contributions disappear at about 100\,K, where the Seebeck coefficient shows a slight upturn. The rather small magnitude of the thermopower for larger hole dopings is due to a pronounced compensation of electron vs. hole carriers. Note that the very flat Ni-$d_{x^2-y^2}$ dispersion with low QP weight yields negligible contribution for all considered dopings.

Similar to the Seebeck coefficient, the Hall coefficient for the stoichiometric undoped case is determined by the SD electron pockets and therefore negative and much larger than that for the higher hole dopings. Be aware that it is scaled by a factor of 1/20 to properly show it within Fig.\ref{Figure1}(b). For dopings $x=0.16$ and $x=0.3$ the intriguing interplay between the SD band around A and the relevant quite flat Ni-$d_{z^2}$ branch makes the understanding of the definite evolution of the Hall coefficient rather tricky. Correspondingly, the value is much smaller, yet for x=0.3 a robust sign change is revealed at $T\sim 170$\,K. This qualitative behavior agrees with experimental Hall data~\cite{li20,zeng20}.

In general, at stoichiometry both the Seebeck and the Hall coefficients display negative and therefore, from an idealistic single-band perspective, dominant electron-like transport. With hole doping this character is
initially diminishing in favor of a stronger hole-like component. Interestingly for $x=0.16$, a weakly hole-like Seebeck and weakly electron-like Hall coefficient are revealed, rendering this
(superconducting) doping regime of intriguing multi-band nature. This is 
in line with the QP dispersion analysis performed in Ref.~\onlinecite{lechermann2020multiorbital}. And again, the small Hall coefficient for larger $x$ results from nearly compensating electrons and 
holes. Note that a small thermopower/Hall response may also generally be realized by a strongly enhanced concentration of carriers, which could
more directly here relate to the itinerant flat Ni-$d_{z^2}$ band. Such a less pronounced role of the SD pocket at the A point may be more fitting to the full
DFT+sicDMFT spectral function with hole doping~\cite{lechermann21}.

The optical conductivity, presented in Fig.~\ref{Figure2} is not restricted to contributions from the proximity of the Fermi level and allows therefore deeper insights into the energy dependence of the electronic excitations. Besides a sharp Drude peak at $\omega = 0$ all the corresponding curves in Fig.~\ref{Figure2} show a second peak followed by a plateau towards higher energies.  As before, the Ni-$d_{x^2-y^2}$-band is of negligible importance and the features stem from excitations between Ni-$d_{z^2}$ and SD band. A peak is located at energies slightly higher than the energy difference of the effective bands at the $\Gamma$ point. In its proximity the dispersion of both bands are quite similar up to a constant energy shift. This results in good vertical nesting conditions which give rise to that peak. The peak-plateau structure is shifted to lower energies for higher hole dopings $x$, which is mainly connected to the $x$-dependent upward shifting of the Ni-$d_{z^2}$ band.
Note that in a very first experimental optics study of Nd$_{0.8}$Sr$_{0.2}$NiO$_2$ by Cervasio {\sl et al.}, a sizable Drude peak and not too strong electron correlations were reported in the normal state~\cite{cervasio22}. This could be in line with the rather itinerant Ni-$d_{z^2}$ band character in this region.

\section{Discussion and Conclusion}
Using a realistic low-energy model for infinite layer nickelates, we find good qualitative agreement with measured Hall data for stoichiometry and hole dopings $x= 0.16,0.3$. A temperature-dependent sign change from negative to positive values upon lowering $T$ happens in our modeling at larger $x$, whilst the corresponding experimental sign change takes place for $x\sim 0.2$~\cite{li20,zeng20}. Agreeing with first experimental suggestions~\cite{li19} this sign change is due to multiband effects. However, it is not due to a competition between the hole pockets of the Ni-$d_{x^2-y^2}$-band and the electron pockets of the SD band as suggested in Ref.~\onlinecite{zeng20}. In our theory, the former band have negligible contribution because of its flatness and small QP weight. Instead, the Fermi surface is reconstructed as the Ni-$d_{z^2}$ band shifts upwards within the $k_z=1/2$ plane of the Brillouin zone with doping $x$. Avoided crossings near the Fermi level result in a more complicated structure in which the SD band is no longer contributing as expected from a straightforward electron pocket. Both, SD band and Ni-$d_{z^2}$ band and their interplay are thus of importance to obtain the sign change.  

To further appreciate the role of the Ni-$d_{z^2}$ band, we showed theoretical results for the Seebeck coefficient. Essentially, we expect it to also change the sign from negative to positive with increasing hole doping. In our model this happens due to low-energy band reconstructions as the Ni-$d_{z^2}$ dispersion crosses $\varepsilon_{\rm F}$, whereas contributions of the effectively localized Ni-$d_{x^2-y^2}$ state are again rather small. A recent experimental assessment of the thermopower in the multi-layer low-valence nickelates Nd$_4$Ni$_3$O$_8$ and superconducting Nd$_6$Ni$_5$O$_{12}$ by Grissonnanche {\sl et al.}~\cite{grissonnanche2022} yield a small, nearly $T$-independent absolute value $<0.1\,\mu$V/K$^2$ and a negative sign for $S$. These multi-layer systems are formally in the well-overdoped
region of infinite-layer materials, and thus the small magnitude merged with weak temperature dependence matches qualitatively with our results. The different sign for this small-magnitude regime may be accidental, however also note that the low-energy region of the multilayer nickelates with multi-Ni-site unit cells hosts enlarged complexity compared to the infinite-layer one~\cite{lechermann22}.

To complete the study, we also considered the optical conductivity. A characteristic peak-plateau structure is observed, which is connected to excitation between Ni-$d_{z^2}$ band and SD band, which is shifted to lower energies with doping. Comparing the effective QP band structure in Fig.~\ref{Figure3} with the more general DFT+sicDMFT spectral function  \cite{lechermann2020multiorbital}, our model does clearly not contain all contributions in the relevant energy range. Note e.g. that the low-lying Ni-$t_{2g}$ states are excluded in the three-band description.
However, the striking features of our model study should still be visible in the full description.  For instance, we expect the peak-plateau structure to be visible in optics measurements and observations of its shift to lower energies could provide evidence for the doping dependent shift of the Ni-$d_{z^2}$ band.

In conclusion, we here provided a quasiparticle assessment of the transport in infinite-layer nickelates. The qualitative agreement with existing experimental data for the doping-dependent Hall coefficient is encouraging and we look forward to such comparisons between data from our predictions and future transport measurements. As noted, we do not expect a QP-based description to be fully conclusive for these challenging nickelates, but it serves as an important reference for further investigations of transport in these materials.

\section{Acknowledgements} 

The work is supported by the German Research Foundation within the bilateral NSFC-DFG Project ER 463/14-1. 

 \bibliography{literatur}
\end{document}